\documentclass[11pt]{article} % use larger type; default would be 10pt

\usepackage[utf8]{inputenc} % set input encoding (not needed with XeLaTeX)

\usepackage{geometry} % to change the page dimensions
\usepackage{graphicx} % support the \includegraphics command and options
\usepackage{booktabs} % for much better looking tables
\usepackage{amsmath}
\usepackage{bm}
\usepackage{hyperref}
\usepackage{mathtools}
\usepackage{array} % for better arrays (eg matrices) in maths
\usepackage{paralist} % very flexible & customisable lists (eg. enumerate/itemize, etc.)
\usepackage{verbatim} % adds environment for commenting out blocks of text & for better verbatim
\usepackage{subfig} % make it possible to include more than one captioned figure/table in a single float
\usepackage[numbers, sort&compress]{natbib}
\usepackage{fancyhdr} % This should be set AFTER setting up the page geometry
\usepackage{sectsty}
\allsectionsfont{\sffamily\mdseries\upshape} % (See the fntguide.pdf for font help)
\usepackage[nottoc,notlof,notlot]{tocbibind} % Put the bibliography in the ToC
\usepackage[titles,subfigure]{tocloft} % Alter the style of the Table of Contents

\usepackage{xcolor} % for \textcolor
\title{Approximate Macroscopic Dynamics of Spiking Neural Networks Based on Solutions to the Transport Equation}

\author{Wilten Nicola \textsuperscript{1,2,3*} and Sue Ann Campbell\textsuperscript{4,5}
\\
$^\text{1}$ Hotchkiss Brain Institute, 
\\
$^\text{2}$  Department of Cell Biology and Anatomy, University of Calgary, Alberta, Canada
\\
$^\text{3}$  Department of Physics and Astronomy, University of Calgary, Alberta, Canada
\\
$^\text{4}$  Department of Applied Mathematics, University of Waterloo, Ontario, Canada
\\
$^\text{5}$ Centre for Theoretical Neuroscience, University of Waterloo, Ontario, Canada
\\
$^\text{*}$ Corresponding author: wilten.nicola@ucalgary.ca}

\begin{document}
\maketitle
%\linenumbers
\begin{abstract}
Firing rate fluctuations in neural populations are observed experimentally over multiple time scales, in single neurons, across trials when elicited by stimuli, and across populations.  In this work, we examine how firing rate fluctuations emerge in networks of coupled integrate-and-fire neurons as a function of the initial distribution of voltages in networks with time-varying inputs.  We analytically derive an approximation for the evolution of the instantaneous population rate or \emph{flux} as a function of the initial voltage distribution through a Fokker-Planck system.  Unlike earlier mean field approaches based on asynchronous or constant flux steady state solutions to the Fokker-Planck system, the approach considered here is based on the transport solution to the advection equation and assumes that the time-varying inputs are slow, and the neurons are in the excitation driven regime.  The transport mean field system predicts how firing rate fluctuations emerge from a dynamic interaction between time-varying inputs, initial densities, and coupling in populations of neurons. 
\end{abstract}

\section*{\textbf{Introduction}} 

Networks of neurons display fluctuations in their population activities at short time scales \cite{gamma1}, long time scales \cite{pop1}, induced by stimuli \cite{mehrdad,stim1}, and across trials \cite{mehrdad}. A theory of how these fluctuations are linked to the underlying neuronal voltages, the network connectivity, and time-varying inputs has remained elusive. 

However, considerable progress has been made in describing the collective dynamics and behaviours of neurons with mean field theories, where the dynamics of the entire population of neurons are reduced to systems of equations for key average quantities (or higher order moments) that qualitatively and quantitatively describe the network's behaviour \cite{nesse,nicola_2d,nicola_het,nicola_noise,montbrio2,gast1,gast2,bi2021asynchronous,chen2,brunel1,brunel2,ostojic1}.  Early mean field theories were phenomenological \cite{cowan1}, but provided qualitative insight into the emergence of different dynamical states for populations of neurons \cite{wilsoncowan1,wilsoncowan2,wilsoncowan3,wilsoncowan4,wilsoncowan5}.  Later mean field theories were derived directly from networks of integrate-and-fire neurons through the Fokker-Planck equation \cite{nesse,nykamp1,nicola_het,nicola_analysis}.  Typically, these systems were derived under the assumption that the asynchronous voltage distribution of the population was stable \cite{cvv1,nicola_analysis,nicola_2d}.  While these systems could predict the qualitative behaviours of the networks under consideration, they were often unable to predict fast fluctuations of the instantaneous population activity. 

More recently, modern mean field or neural mass models that can resolve firing rate fluctuations have been developed \cite{luke2013complete,chen1,chen2,montbrio1,montbrio2,montbrio3,montbrio4,devalle2018dynamics,gast2020mean,gast2021mean,gast1,gast2,coombes}.  These mean field systems assume heterogeneous networks based on the Ott-Antonsen ansatz \cite{ott1}, where neuronal currents and voltage distributions are Lorentzian (Cauchy).  This assumption on the underlying neural heterogeneity allows for the derivation and subsequent analysis of low-dimensional mean field systems that couple the firing rate and the mean-voltage together \cite{chen1,chen2,gast2020mean,gast2021mean,gast2}.  Further, modern neural mass models are also analytically exact for a simulated network, with the exception of finite-size effects and other sources of numerical imprecision.  The only disadvantages to modern neural mass models are that they are not readily applied to all neuron models, being critically dependent on the quadratic integrate-and-fire model or derivatives of it \cite{chen1,chen2,montbrio1}. Other approaches solve for the first few dominant eigenmodes of the Fokker-Planck operator numerically and use these to predict the fluctuations of firing rates across networks \cite{ostojic1}. 

Here, we consider an alternative formulation for mean field theories based on solutions to the transport equation \cite{risken}. These solutions emerge as approximations in weakly coupled networks or networks with slowly varying inputs.  We derive the transport solution to a Fokker-Planck network of 1-dimensional integrate-and-fire neurons.  This yields an operator which reversibly transforms initial voltage densities to time-dependent population firing rates. This time varying firing rate is more commonly referred to as the flux in Fokker-Planck theory.  We show that the transport solution can be used as the basis for constructing a transport based mean field system for a coupled network of neurons with time-dependent inputs assuming that the inputs are slow, and (or) the coupling is weak.  The transport mean field system is able to capture the time varying fluctuations of the firing rate as a function of both time-varying inputs and connectivity when the neurons are operating in the excitation driven regime, for any one-dimensional integrate-and-fire neuron type. 

\section*{\textbf{Results}} 

\subsection*{\textbf{Background}}

We consider networks of all-to-all coupled integrate-and-fire neurons with inputs: 
\begin{eqnarray}
\frac{dv_i}{dt} = F(v_i) + I + ws(t) + c(t) \\
v_i(t^-) = v_{thr}\rightarrow v_i(t^+) = v_{reset} 
\end{eqnarray}
where $F(v)$ governs the single neuron subthreshold dynamics, $c(t)$ acts as an external input, and $w$ weights the global coupling variable $s(t)$.  The coupling variable $s(t)$ acts as a filter for the spikes:
\begin{eqnarray}
\tau_s \frac{ds}{dt} = -s +\frac{1}{N}\sum_{j=1}^N \sum_{t<t_{jk}}\delta(t-t_{jk}) \label{sde}
\end{eqnarray}
where $t_{jk}$ is the $k$th spike fired by the $j$th neuron and $\tau_s$ is the filtering time constant while $\delta$ is the Dirac delta. Note that for simplicity, the membrane time constant has been absorbed into the systems time variable.  While our derivation applies to any $F(v)$, we will primarily consider the leaky ($F(v) = -v$) and quadratic ($F(v)= v^2$) integrate-and-fire neurons as exact solutions can be derived in these cases.  We will also primarily consider the case where $F(v) + I + ws(t) + c(t)>0$ for all $t>0$, that is, the neurons are operating in the excitation driven regime.  We outline in the discussion the steps required when this is not the case. 

To simplify the derivation considerably, define $\bar{I}(t) = I + ws(t) + c(t)$ and consider the network given by 
\begin{eqnarray}
\frac{dv_i}{dt} &=& F(v_i) + \bar{I}(t) \label{veq}\\
v_i(t^-) &=& v_{thr}\rightarrow v_i(t^+) = v_{reset}. \label{reset}
\end{eqnarray}
To study the dynamics at the population level we define the population density function $\rho^v(v,t)$ as the density of neurons at a point $v$ in phase space 
 at time t. In the large network limit $N\rightarrow\infty$, the principle of conservation of probability leads to a Fokker-Planck partial differential equation (PDE) for the population density function, supplemented by the ordinary differential equation (ODE) for the synaptic filter:  %that is, the continuity equation,
%At the population level, the system is governed by a Fokker-Planck partial differential equation given by:
\begin{eqnarray}
\frac{\partial \rho^v(v,t)}{\partial t} &=& -\frac{\partial}{\partial v}\left(\rho^v(v,t)(F(v) + \bar{I}(t)\right) = -\frac{\partial J^v(v,t)}{\partial v}\label{fpe1}\\
\tau_s \frac{ds}{dt} &=& -s + J^v(v_{thr},t) \\
\rho^v(v,0) &=& \rho_0^v(v) \label{fpe2} \\
J^v(v_{thr},t) &=& J^v(v_{reset},t) \label{BC}
 \end{eqnarray}
 Here $\rho_0^v(v)$ corresponds to the initial voltage density at time $t=0$ and the boundary condition~\eqref{BC}, comes from the reset equation~\eqref{reset} in the single neuron model. The quantity $J^v(v,t)$ represents the flux of neurons through the point $(v,t)$ induced by the single neuron ODE~\eqref{veq}. As shown in \cite[Appendix 1.1]{nicola_2d}, in the limit $N\rightarrow\infty$ the  source term in the synaptic filtering equation ~\eqref{sde} converges to the instantaneous network averaged firing rate, which is equal to the flux through the voltage threshold, $J^v(v_{thr},t)$.

 In prior mean field or neural mass analyses \cite{cvv1,nesse,nicola_2d,nicola_analysis,nicola_het,nicola_noise}, the Fokker-Planck system in equations (\ref{fpe1})-(\ref{fpe2}) consisting of a coupled PDE/ODE pair is transformed into a single non-smooth ODE by assuming that the synaptic filtering $s(t)$ and external input $c(t)$ are slow. Hence the partial differential equation for $\rho^v(v,t)$ equilibrates quickly to the asynchronous density ($\rho^v(v,t) = \rho^v_{AS}(v)$) where 
 \begin{eqnarray}
 \rho^v_{AS}(v) = \frac{\Omega(\bar{I}(t))}{F(v) + \bar{I}(t)} \label{asy1}
 \end{eqnarray}
 and $\Omega(\bar{I}(t))$ is the single neuron oscillation rate as a function of the slow variable $\bar{I}(t)$.  
 \begin{eqnarray}
\Omega(\bar{I}(t)) = \begin{cases} \left(\int_{v_{reset}}^{v_{thr}}\frac{dv'}{F(v')+\bar{I}(t)}\right)^{-1} & F(v) + \bar{I}(t)>0, \forall v\in [v_{reset},v_{thr}]\\
0 & \text{otherwise}
\end{cases}
 \end{eqnarray}
 leading to the single non-smooth ODE: 
 \begin{eqnarray}
\tau_s s' = -s + \Omega(\bar{I}(t)) = -s + \Omega(I + ws(t) + c(t)) \label{as1}
 \end{eqnarray}
 We refer to equation (\ref{as1}) as the asynchronous mean field system.  Note that the density is asynchronous in the sense that the flux becomes constant, and in the phase domain, the neurons are equally distributed on a circle moving at a constant phase velocity given by the flux \cite{cvv1}. Here, we show that the asynchronous density $\rho_{AS}^v(v)$ is not the only asymptotic density solution one can consider for $\rho(v,t)$.   We explicitly derive an alternate mean field system based on transport solutions of the advection equation, and we refer to this system as the transport mean field system.  The transport mean field system is given by the equations: 
 \begin{eqnarray}
\tau_s \frac{ds}{dt} &=& -s + \nu(t) \label{mf1}\\
\nu(t) &=& \rho_0^v(\tilde{v}(t))(F(\tilde{v}(t))+\bar{I}(t))\\
\tilde{v}(t) &=& g^{-1}\left(\text{mod}\left(-\int_0^t \Omega(\bar{I}(t)),1\right)\right)\\
g(v) &=& \int_{v_{reset}}^v \frac{\Omega(\bar{I}(t))}{F(v)+\bar{I}(t)} \label{mf2}\\
\bar{I}(t) &=& I + ws(t) + c(t)
 \end{eqnarray}
where $\text{mod}(x,1)=x-\lfloor x\rfloor$.  The function $g(v)$ is the probability integral transform for $v$ \cite{cvv1}, which also coincides with a neurons' phase.  The transport mean field system also approximates the time evolution of the voltage density with:
\begin{eqnarray}
\rho(v,t) &=& \rho_0^v(Q(v,t))\frac{F(Q(v,t))+\bar{I}(t)}{F(v) +\bar{I}(t)}\label{mf4}\\
Q(v,t) &=& g^{-1}\left(\text{mod}\left(g(v)-\int_0^t \Omega(\bar{I}(t)),1)\right)\right) 
\end{eqnarray}
although solving for the time evolution of the density is not explicitly required for computing the time evolution of the flux $\nu(t)$, and therefore $s(t)$.

Before deriving the transport mean field equations (\ref{mf1})-(\ref{mf2}) and (\ref{mf4}), we highlight the differences and similarities between the transport mean field system based on equations (\ref{mf1})-(\ref{mf2}) and the asynchronous mean field system based on equation (\ref{as1}).  First, we remark that  in the derivations of both the asynchronous mean field and transport mean field systems, the dynamics of $s(t)$ and $c(t)$ are assumed to be slow. For the transport system, however, the assumption on $s$ can be replaced by an assumption of weak coupling. %also suffices as a condition for the transport mean field system to be a good approximation.  
Further, both the asynchronous and transport mean field systems are closed systems of equations, and can be readily simulated without simulating a full network.  Finally, we highlight the most important difference: the transport mean field system explicitly links the initial density to the firing rate fluctuations via $\nu(t)=J(v_{thr},t)=\rho^v_0(\tilde{v}(t))(F(\tilde{v}(t))+\bar{I}(t))$.  While equations (\ref{mf1})-(\ref{mf4}) seem formidable, they are easily applied when $\rho_0^v(v)$ is known and when $g(v)$ and its inverse are analytically defined, as in the case of the leaky and quadratic integrate-and-fire neurons which we show below.

\subsection*{\textbf{Derivation of the Transport Mean Field Equations}}

Consider the network equations again as:
\begin{eqnarray}
\frac{dv_i}{dt} = F(v_i) + \bar{I}(t), \quad i =1,2,\ldots N
\end{eqnarray}
and consider the probability-integral-transform \cite{cvv1}:
\begin{eqnarray}
u =g(v)=\Omega(\bar{I}(t))\int_{v_{reset}}^{v}\frac{dv'}{F(v')+\bar{I}(t)}\label{tform1}
\end{eqnarray}
where $\Omega(\bar{I}(t))$ is the instantaneous oscillation frequency of a single neuron receiving the current $\bar{I}(t)$:
\begin{eqnarray}
\Omega(\bar{I}(t)) = \left(\int_{v_{reset}}^{v_{thr}}\frac{dv'}{F(v')+\bar{I}(t)}\right)^{-1}
\end{eqnarray}
Note that the function $g(v)$ is also a function of time through $\bar{I}(t)$.  However, we have suppressed the explicit time dependence for convenience.

The function $g(v_i)$ transforms the integrate-and-fire neuron with dynamics $\dot{v}_i =F(v_i)+\bar{I}(t)$ and voltage reset/peak at $[v_{reset},v_{thr}]$ into:
\begin{eqnarray}
\frac{d u_i}{dt}= \Omega(\bar{I}(t)) + \frac{\partial g}{\partial \bar{I}} \frac{d\bar{I}}{dt}\quad i=1,2\ldots N
\end{eqnarray}
Applying the transform to the reset equation~\eqref{reset}, shows that
the new ``voltage" threshold and reset are at $u_{thr}=1$ and 
$u_{reset}=0$, respectively. 
The variable $u_i(t)$ effectively acts as the phase for the $i^{th}$ neuron. 
 
Now we assume that $c$ acts on a slow timescale relative to the individual neuron dynamics. Hence we write $c=c(\epsilon t)$ where $0<\epsilon\ll\Omega(\bar{I}(t))$. This gives
\begin{eqnarray*}
\frac{d\bar{I}}{dt}&=&w\frac{ds}{dt}+\frac{dc}{dt}
\end{eqnarray*}
Thus if $c$ is slow and either the coupling is weak, $w=\mathcal{O}(\epsilon)$, or $s$ is slow,
$\tau_s=\mathcal{O}(\frac{1}{\epsilon})$, then $\bar{I}(t)$ is slow
\begin{eqnarray}
\frac{d\bar{I}(t)}{dt}=\mathcal{O}(\epsilon)\ll \Omega(\bar{I}(t)).
\end{eqnarray}
%we assume that $\bar{I}(t)$ is slow: 
%\begin{eqnarray}
%\frac{d\bar{I}(t)}{dt}\ll \Omega(\bar{I}(t))
%\end{eqnarray}
%which 
This implies that the phase dynamics are approximately determined by 
\begin{eqnarray}
\frac{du_i}{dt} \approx \Omega(\bar{I}(t)).
\end{eqnarray}

In the $u$-domain, the $O(1)$ equations to the Fokker-Planck equation are given by:
\begin{eqnarray}
\frac{\partial \rho^u(u,t)}{\partial t} &=& -\Omega(\bar{I}(t))\frac{\partial \rho^u(u,t)}{\partial u} \label{pdeu}\\
J^u(u,t) &=&\Omega(\bar{I}(t)) \rho^u(u,t)\\
\rho^u(0,t) &=& \rho_0^u(u) \label{pdeu2}
\end{eqnarray}
with the boundary condition 
\begin{eqnarray}
J^u(1,t) = J^u(0,t) \label{pdeu3}
\end{eqnarray} 
where $\rho_0^u(u)$ is given by 
\begin{eqnarray}
\rho_0^u(u) = \rho_0^v(g^{-1}(u))\left|\frac{dg^{-1}}{du}\right|.
\end{eqnarray}

The system of equations (\ref{pdeu})-(\ref{pdeu2}) with the boundary condition (\ref{pdeu3}) has an exact analytical solution:
\begin{eqnarray}
\rho^u(u,t) = \rho_0^u\left(\text{mod}\left(u - \int_0^t \Omega(\bar{I}(t')dt'),1)\right)\right)\label{rhou}
\end{eqnarray}
where the modulus function enforces the periodic boundary condition:
\begin{eqnarray}
\mod(x,1) = x - \lfloor x\rfloor.
\end{eqnarray}
With the solution (\ref{rhou}), the density in $v$ as function of time becomes: 
\begin{eqnarray*}
\rho^v(v,t) &=&  \rho^v_0\left(g^{-1}\left(\text{mod}\left(g(v) -  \int_0^t \Omega(\bar{I}(t')dt'),1\right)\right)\right) \frac{g'(v)}{g'(g^{-1}(\text{mod}(g(v)- \int_0^t \Omega(\bar{I}(t')dt'),1)))} \\
&=& \rho^v_0\left(Q(v,t)\right) \left(\frac{F\left(Q(v,t)\right)+\bar{I}(t)}{F(v)+\bar{I}(t)}\right)\\
Q(v,t) &=& g^{-1}\left(\text{mod}\left(g(v) -  \int_0^t \Omega(\bar{I}(t')dt'),1\right)\right)
\label{rho1} 
\end{eqnarray*}
while the flux is given by 
\begin{eqnarray*}
J(v,t) &=&(F(v)+I)\rho_v(v,t) \\
&=&  \rho^v_0\left(g^{-1}\left(\text{mod}\left(g(v) -  \int_0^t \Omega(\bar{I}(t')dt'),1\right)\right)\right) \frac{1}{g'(g^{-1}(\text{mod}(g(v) -  \int_0^t \Omega(\bar{I}(t')dt',1)))} \\
&=& \rho^v_0\left(Q(v,t)\right) \left[F\left(Q(v,t)\right)+\bar{I}(t)\right]
\end{eqnarray*}
This solution immediately yields the time varying network averaged firing rate:
\begin{eqnarray}
%\nu(t) = J(v_{thr},t)=\rho^v_0(g^{-1}(\text{mod}(- \int_0^t \Omega(\bar{I}(t')dt',1))) \left[F\left(\left(g^{-1}\left(\text{mod}(- \int_0^t \Omega(\bar{I}(t'))dt',1\right)\right) \right)+\bar{I}(t)\right]  \label{nu1}
\nu(t) = J(v_{thr},t)=\rho^v_0(Q(v_{thr},t)) \left[F\left(Q(v_{thr},t)\right)+\bar{I}(t)\right],  \label{nu1}
\end{eqnarray}
where 
\begin{eqnarray}
Q(v_{thr},t)=g^{-1}\left(\text{mod}\left(- \int_0^t \Omega(\bar{I}(t')dt',1)\right)\right).
\end{eqnarray}
Defining $\tilde{v}(t)=Q(v_{thr},t)$, the firing rate dynamics can be conveniently written as:
\begin{eqnarray}
\nu(t) =\rho^v_0(\tilde{v}(t))\left(F(\tilde{v}(t))+\bar{I}(t)\right),
\end{eqnarray}
where 
%\begin{eqnarray}
%\tilde{v}(t) = Q(v_{thr},t)=g^{-1}\left(\text{mod}\left(- \int_0^t \Omega(\bar{I}(t')dt',1)\right)\right)
%\end{eqnarray}
%and 
$\tilde{v}$ is a proper voltage through the application of $g^{-1}$ to the interior argument.

\subsection*{\textbf{Single, Uncoupled Population with Constant Input}}
\begin{figure}[htp!]
\centering
\includegraphics[scale=0.65]{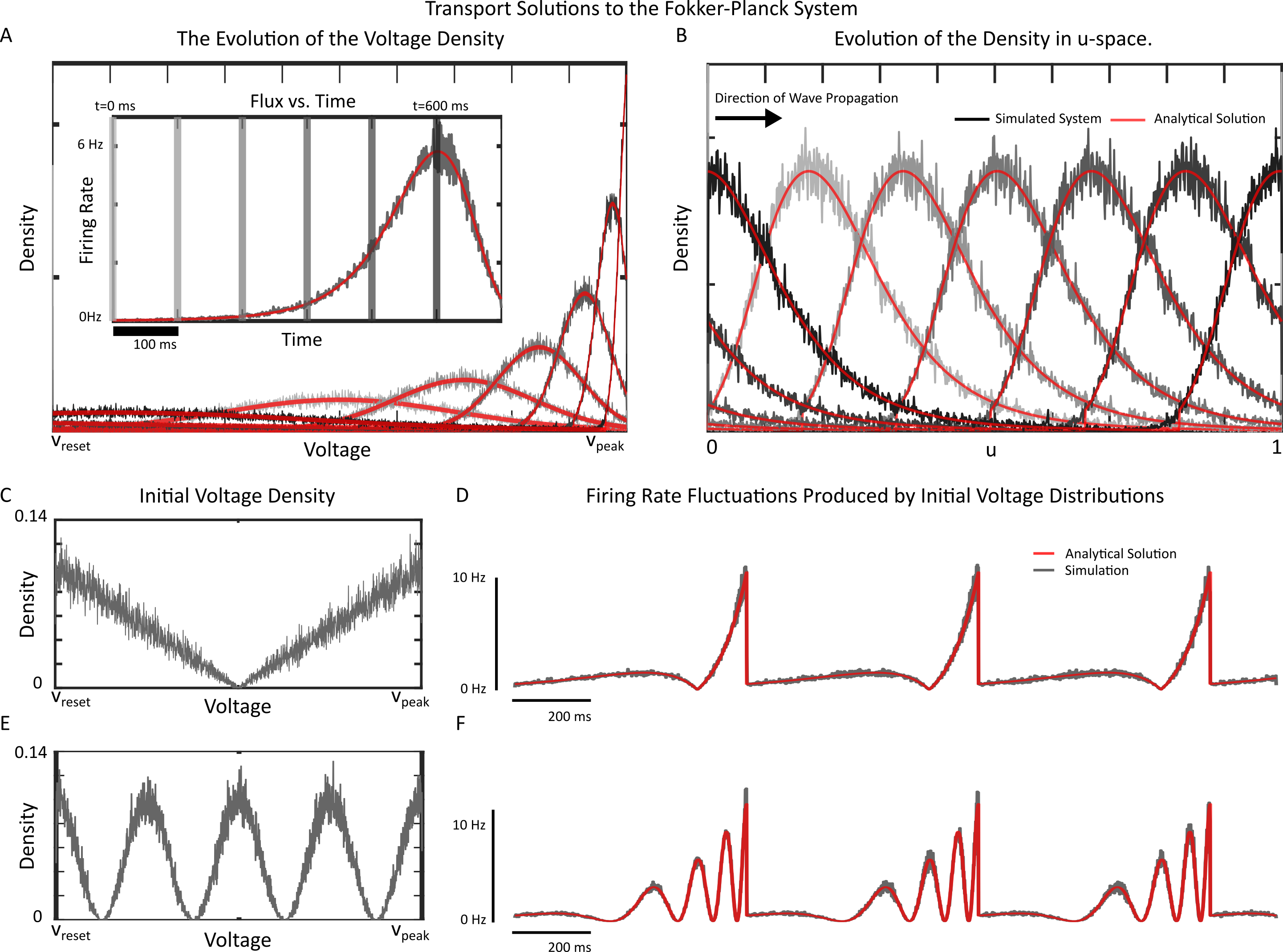}
\caption{\textbf{The Firing Rate Fluctuations caused by the Initial Voltage Density can be Predicted in Populations of Integrate-and-Fire Neurons} \textbf{(A)} The flux, or the population firing rate $\nu(t)$ as a function of time (inset), and the population density of an uncoupled network of leaky integrate-and-fire neurons at different time points.   The vertical lines in the inset correspond to the times at which the voltage density was sampled.  
\textbf{(B)}  By applying the probability integral-transform, the voltages $v$ can be transformed with $v=g(u)$ to a space where the Fokker-Planck equation has an easily solvable transport solution as a function of the initial density (also transformed to $u$).   The flux, $\nu(t)$ can be predicted as a function of the initial voltage, $\rho_0(v)$ with this substitution. 
\textbf{(C)}  The initial voltage density given by $\rho_0(v)=|v-\frac{v_{reset}+v_{thr}}{2}|$.   
\textbf{(D)}  The predicted (red) versus empirically estimated firing rate fluctuations (grey) $\nu(t)$ for the density in (C) with a network of $N=10^5$ LIF neurons.  
\textbf{(E)}  The initial voltage density given by $\rho_0(v)=\kappa(1+\sin(2\pi v))$ where $\kappa$ is a normalization constant over the interval [$v_{reset},v_{thr}$]
\textbf{(F)} The predicted (red) versus empirically estimated firing rate fluctuations (grey) $\nu(t)$ for the density in (E) with a network of $N=10^5$ LIF neurons
}\label{figure3}
\end{figure}

Before considering coupled populations of neurons further, we remark that when $\bar{I}(t)=I$ is constant and $F(v) + I >0$ for $v\in[v_{reset},v_{thr}]$, the transform into the phase variable and resulting solution of the phase density are simpler to understand (Figure \ref{figure3}A).  In particular, the transform 
$$ u = g(v) = \int_{v_{reset}}^v\frac{\Omega(I)dv'}{F(v')+I}$$
yields a density solution in the $u$(phase)-domain as:
$$\rho(u,t)=\rho^u(\text{mod}(u - \Omega t,1))$$
which corresponds to a density moving at constant velocity $\Omega$ along a ring (Figure \ref{figure3}B).   When converting back to the voltage domain, the flux becomes non-constant leading to distortions throughout the $v$-domain.  

For the leaky integrate-and-fire ($F(v)=-v$) and quadratic ($F(v)=v^2$) cases, the corresponding probability integral transforms ($g(v)$) are
\begin{eqnarray}
g(v)&=& \Omega(I)\left(\log(-v_{reset}+I) - \log(-v+I) \right), \quad (\text{LIF})\\
g(v) &=&\Omega(I)\left(\frac{ \arctan\left(\frac{v}{\sqrt{I}}\right) - \arctan\left(\frac{v_{reset}}{\sqrt{I}}\right)}{\sqrt{I}}\right), \quad (\text{QIF})
\end{eqnarray}
with inverses:
\begin{eqnarray}
g^{-1}(u) &=& I - \exp\left(\frac{u}{\Omega(I)}\right)(-v_{reset}+I),\quad \text{(LIF)},\\
g^{-1}(u) &=& \sqrt{I}\tan\left(\frac{u\sqrt{I}}{\Omega(I)}+\arctan\left(\frac{v_{reset}}{\sqrt{I}}\right)\right), \quad \text{(QIF)}.
\end{eqnarray}
This yields the following firing rate fluctuations for the LIF neuron:
\begin{eqnarray}
\nu_{LIF}(t)&=&  %\rho_0^v(Q(t))\frac{1}{-Q(t)+I}\\
\rho_0^v(\tilde{v}(t))\left(-\tilde{v}(t)+I\right)\\
\tilde{v}(t) &=& I - \exp\left(\frac{\mod(-\Omega(I)t,1)}{\Omega(I)}\right)(-v_{reset}+I)\\
\Omega(I) &=& \left(\log(-v_{reset}+I)-\log(v_{thr}+I)\right)^{-1}
\end{eqnarray}
while for the QIF neuron, the firing-rate fluctuations are:  
\begin{eqnarray}
\nu_{QIF}(t)&=&  \rho_0^v(\tilde{v}(t))\left(\tilde{v}(t)^2+I
\right)\\
\tilde{v}(t)&=&\sqrt{I}\tan\left(\frac{-{\rm mod}(-\Omega(I)t,1)\sqrt{I}}{\Omega(I)}+\arctan\left(\frac{v_{reset}}{\sqrt{I}}\right)\right)\\
\Omega(I) &=& \frac{\arctan\left(\frac{v_{thr}}{\sqrt{I}}\right)- \arctan\left(\frac{v_{reset}}{\sqrt{I}}\right)}{\sqrt{I}}
\end{eqnarray}
%Note that for simplicity, the $\tau_m$ in the LIF case has been absorbed into the systems time variable.  

\subsection*{\textbf{Invertibility of the Flux for an Initial Density}}

Before returning to the full
%numerically validating the 
 transport mean field equations, 
we remark that for a constant current $I$, the flux can be interpreted as an operator mapping from the set of functions that act as densities on $[v_{reset},v_{thr}]$ to the set of firing rates that act as densities on $[0,\Omega(I)^{-1}]$:
\begin{eqnarray}
\nu(t) &=&\rho^v_0(\tilde{v}(t))\left(F(\tilde{v})+I\right)\\
&=& R(\rho^v_0)
\end{eqnarray}
where $\tilde{v}(t)=g^{-1}\left({\rm mod}\left(-\Omega(I) t,1\right)\right).$

\begin{figure}[htp!]
\centering
\includegraphics[scale=0.65]{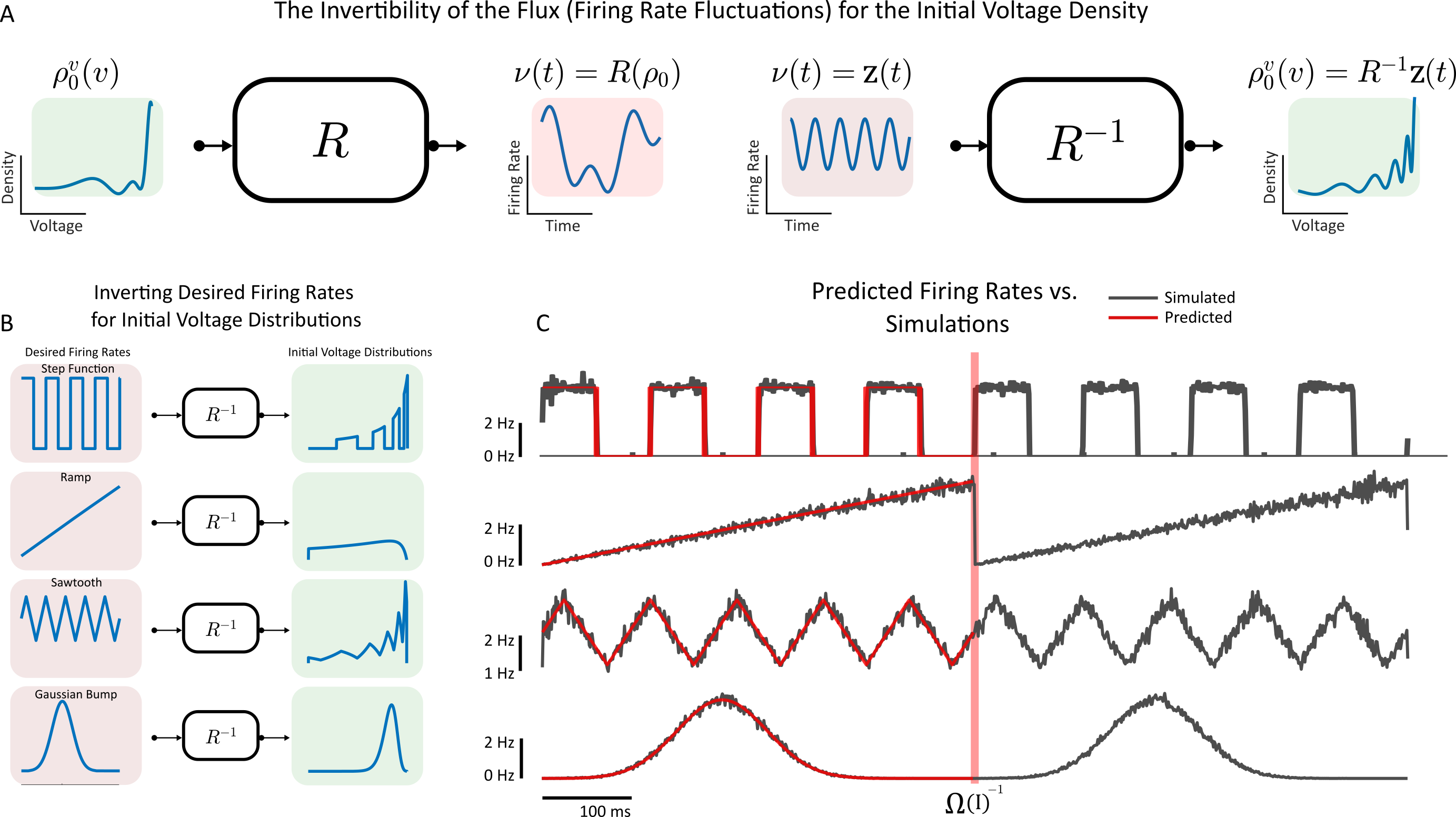}
\caption{\textbf{The Flux Operator is Invertible when Inputs are Constant}
\textbf{(A)} The flux $\nu(t)$ is a function of the initial voltage density $\rho_0^v(v)$, as determined by the operator $R$.  The operator $R$ is invertible, where $R^{-1}$ provides a solution to $R^{-1}z = \rho_0^v(v)$. 
\textbf{(B)}  The desired firing rate distributions on the left vs. derived density (right).   
\textbf{(C)} A simulation of the network initialized with the density $\rho = R^{-1}z$ where $z$ is a square wave (first row), a ramp (second row), a sawtooth wave (third row) and a Gaussian bump (fourth row).  In all cases, $z(t)$ is periodic with a period of $\nu^{-1}$.
}\label{figure4}
\end{figure}

Finally, and somewhat surprisingly, the operator $R$ is explicitly invertible.  Note that $\nu(t)$ only depends on $t$ through $\tilde{v}(t)$.  Suppose we want a particular firing rate $\nu(t)=z(t)$ then the inverse operator is defined by:
\begin{eqnarray}
\rho^v{(\tilde{v})} &=& \frac{z(t^{-1}(\tilde{v}))}{F(\tilde{v})+I}\\
&=& R^{-1}(z) \label{eqfr}\\
  t^{-1}(\tilde{v}) &=& \frac{g(\tilde{v})-1}{\Omega(I)}      
\end{eqnarray}
We verified the invertibility of the flux operator for uncoupled networks in Figure \ref{figure4} (A-C) by solving for the initial densities using equation (\ref{eqfr}) for a square wave flux, a ramping flux, a sawtooth flux, and a gaussian bump (Figure \ref{figure4}B-C)

\subsection*{\textbf{Single-Coupled Population of Neurons}}

\begin{figure}[htp!]
\centering
\includegraphics[scale=0.5]{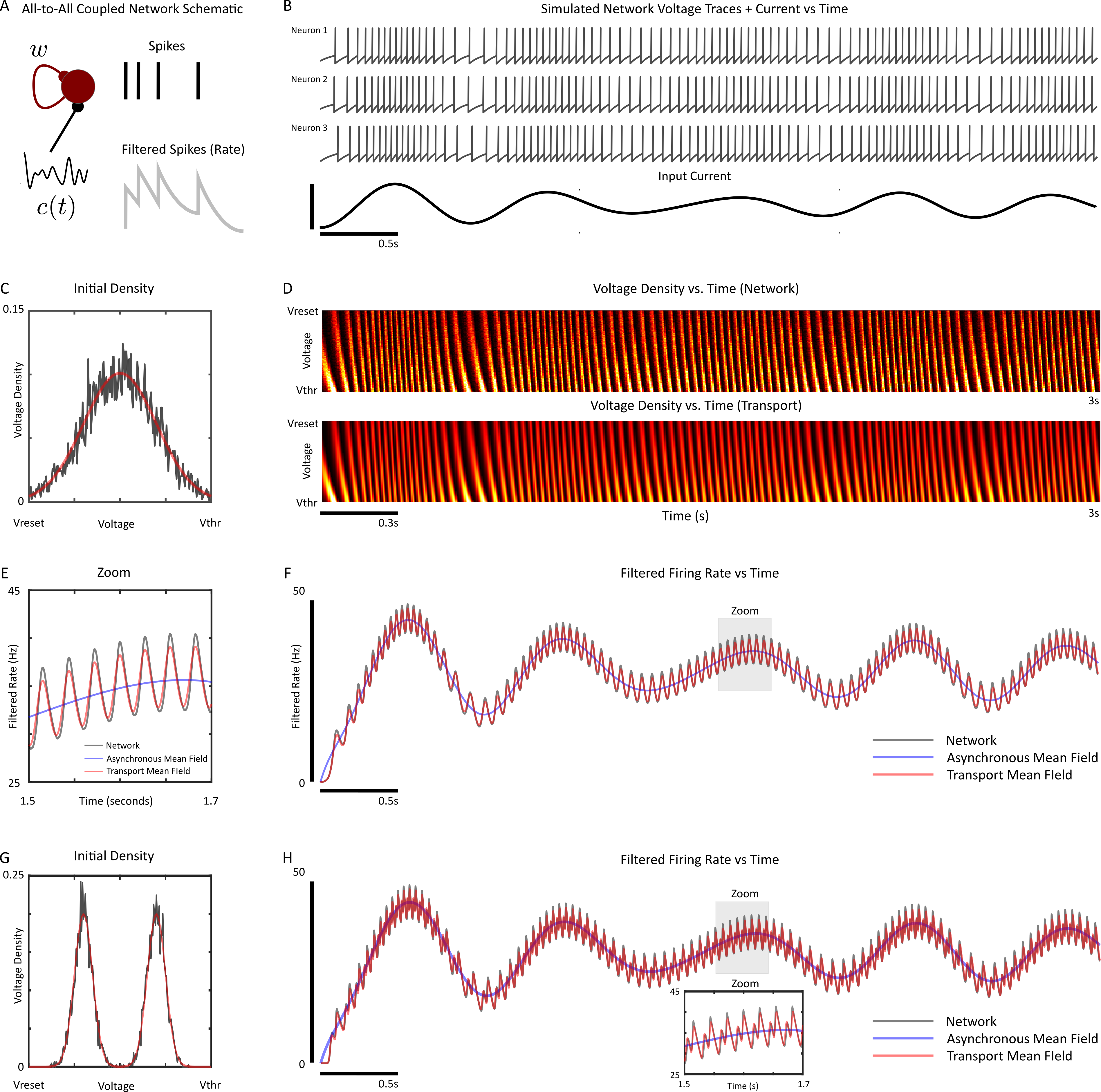}
\caption{\textbf{Mean Field Systems Based on Transport Solutions to the Fokker-Planck System} 
\textbf{(A)} A schematic of an all-to-all coupled network of integrate-and-fire neurons receiving an input $c(t)$.  The neurons are coupled by the global weight $w$ (see main text).  The spikes are filtered with a single-exponential filter. 
\textbf{(B)} The spike raster plot for 3 neurons in a network of $N=5000$, with a slowly varying input current.  
\textbf{(C)}  A Gaussian unimodal initial density with mean $\mu = -50$ mV and standard deviation $\sigma = 3$ mV.  The probability density function is shown in red, while the empirical density is shown in black.   
\textbf{(D)}  A heatmap of the density vs. time for the simulated network (top) vs. the transport mean field approximation (bottom) 
\textbf{(E)} 
A zoom of the filtered firing rates from (F).  
\textbf{(F)}  The filtered firing rates for the network considered in (D).  The simulated network (black) is plotted against the asynchronous mean field (blue) and the transport mean field (red)
\textbf{(G)} A bi-modal initial density with each mode being a Gaussian with a standard deviation of 1 mV, with the means being $-54$ mV and $-46$ mV.  
\textbf{(H)} The filtered firing rates for the simulated network initialized as in (G).  An inset showing a zoom of the firing rate fluctuations is also displayed. }
\label{figure1}
\end{figure}

To test the transport mean field system, we considered a self-coupled network with a time varying input $c(t)$ (Figure \ref{figure1}A).  The network under consideration consisted of leaky-integrate-and-fire neurons ($F(v) = -v$, Materials and Methods, Figure \ref{figure1}B).  In the slowly varying regime, the neuronal spike rate varies with the input current and is amplified/distorted by the coupling strength $w$.  We initialized a network with a unimodal truncated Gaussian initial density in the interval $[v_{reset},v_{thr}]$ (Materials and Methods, Figure \ref{figure1}C).  The empirical voltage density was computed numerically in (Figure \ref{figure1}D) and compared to the transport density solution in equation (\ref{mf4}), which provided a good approximation to the time evolution of $\rho^v(v,t)$. The filtered spike train, $s(t)$ was also well approximated by the transport mean field solution (Figure \ref{figure1}E-F) where the transport mean field solution approximated the fast fluctuations of the firing rate as a function of the initial density, something that the asynchronous mean field could not do. This was also evident when considering more exotic initial density functions, like bimodal densities (Figure \ref{figure1}G-H). 

\begin{figure}[htp!]
\centering
\includegraphics[scale=0.62]{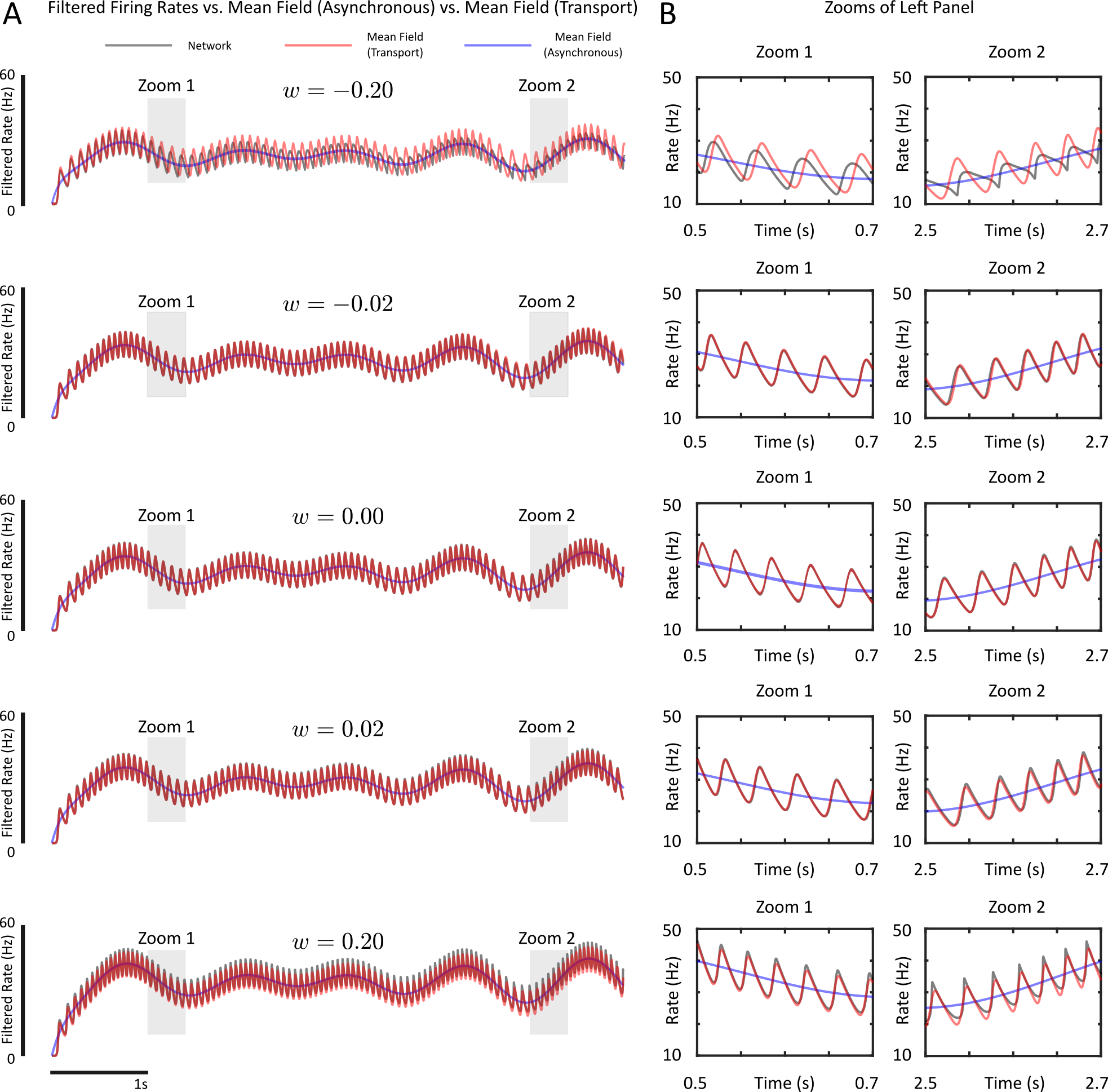}
\caption{ 
Varying the Coupling Strength in Simulated Networks
\textbf{(A)} A simulated network of $N=5000$ Leaky-integrate-and-fire neurons for different levels of coupling strength.  The filtered firing rates for the network are shown in black, while the transport mean field system is plotted in red and the asynchronous mean field system is plotted in blue.  \textbf{(B)}  Two zoomed intervals corresponding to the simulated network in (A) for $t\in[0.5,0.7]$ s and $t\in[2.5,2.7]$ s. 
}\label{figure2}
\end{figure}

  Further, the transport mean field system could also approximate the fast firing rate fluctuations in $s(t)$ for inhibitory coupling ($w<0$) and excitatory coupling $w>0$ for weak and intermediate coupling strengths (Figure \ref{figure2}A-B).  We remark that for more negative values of $w$ lead to periodic bursting which violates the $F(v) + \bar{I}(t)>0$ condition, while more positive $w$ can lead to blow up for the network considered here.   We also remark that while the transport mean field system is accurate initially, the accuracy decreases asymptotically as $t\rightarrow \infty $ (Figure\ref{figure2}B) as the transport mean system cannot predict the long-time scale evolution of the density caused by changes to neuronal synchrony as the system evolves to the asynchronous, synchronous, or partially synchronized states. 
  
  However, these results show that the transport mean field system outperforms the asynchronous mean field system by providing a better approximation to the fast firing rate fluctuations caused by non asynchronous initial densities. 

\subsection*{\textbf{Two-Coupled Populations of Neurons}}

\begin{figure}[htp!]
\centering
\includegraphics[scale=0.5]{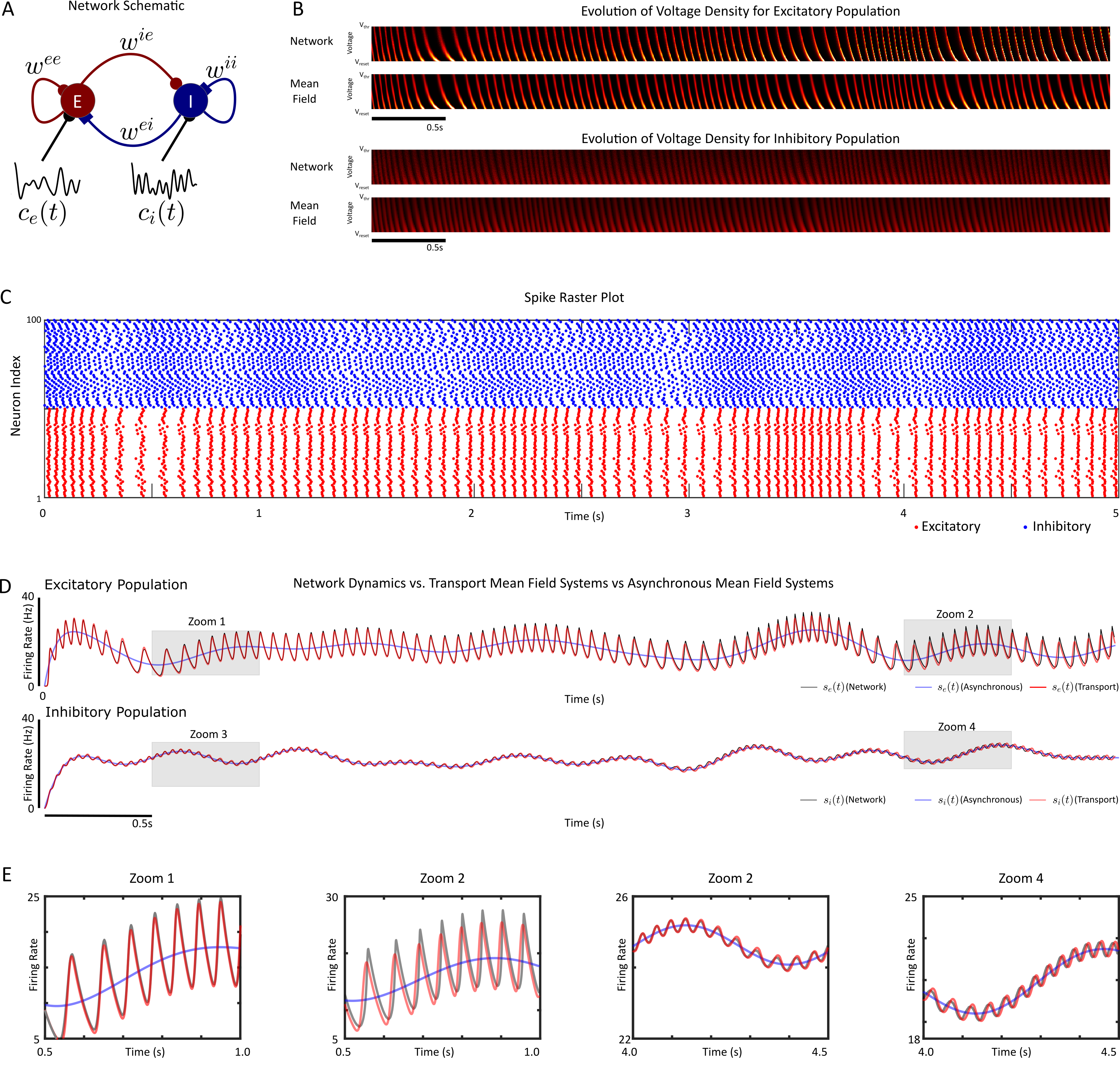}
\caption{\textbf{Mean Field Systems Based on Transport Solutions to the Fokker-Planck System}
\textbf{(A)} A schematic of a coupled population of excitatory (red) and inhibitory (blue) neurons.  The weight $w^{\alpha \beta}$ coupled population $\beta$ to population $\alpha$ where $\alpha/\beta$ = $E/I$.   Each subpopulation receives its own external input $(c_e(t)/c_i(t))$.    
\textbf{(B)} The empirically estimated density for the excitatory population (first row), the solution from the transport mean field equations (second row), the empirical density for the inhibitory population (third row) and the transport mean field equation (fourth row). 
\textbf{(C)} The spike raster plot for $50$ excitatory and $50$ inhibitory neurons.  The total network size was $N^E=N^I=5000$ but only a subset of spikes are plotted for clarity.
\textbf{(D)} The filtered spike rates for the excitatory population (top) and inhibitory population (bottom).  The full network is shown in black, while the transport mean field system is plotted in red, and the asynchronous mean field system in blue. 
\textbf{(E)}
Zooms of the filtered spike rates corresponding to the labeled panels in (D). 
}\label{figure5}
\end{figure}

With the single population numerically studied, 
%and the transport mean field system fully derived, 
we considered networks of coupled and driven excitatory/inhibitory integrate-and-fire neurons (Figure \ref{figure5}A):
\begin{eqnarray}
\frac{dv^E_i}{dt} &=& F(v^E_i) + I^E(t),\quad v_i^E(t^-) = v_{thr}\rightarrow v_i^E(t^+) = v_{reset}, \quad i=1,2,\ldots N^E \label{ve1} \\
\frac{dv^I_i}{dt} &=& F(v^I_i) + I^I(t), \quad v_i^I(t^-) = v_{thr}\rightarrow v_i^I(t^+) = v_{reset} \quad i=1,2,\ldots N^I, \label{vi1}\\
I^E(t) &=& I^E_{bias} + w^{EE}s_E(t)-w^{EI}s_I(t) \\
I^I(t) &=& I^I_{bias} -w^{II}s_I(t) + w^{IE}s_E(t)
\end{eqnarray}
where $F(v)$ governs the neural dynamics for the excitatory/inhibitory population and $w^{\alpha\beta}\geq 0$ for $\alpha,\beta = E/I$.  We will assume that the voltages are drawn from an initial density for each population, $\rho^v_{0,E}(v)$ and $\rho^v_{0,I}(v)$.  As in the single population case, the coupling variables $s_E(t)$ and $s_I(t)$ act as synaptic filters for the spikes of the respective populations:
\begin{eqnarray}
\tau_\alpha\frac{ds^\alpha}{dt} = -s^\alpha + \sum_{j=1}^{N^\alpha}\sum_{t^\alpha_{jk}<t}\delta(t-t^\alpha_{jk}), \quad \alpha = I,E \label{s}
\end{eqnarray}
where $t^\alpha_{jk}$ is the $j$th spike fired by the $k$th neuron in population $\alpha$.  We again primarily consider the leaky-integrate-and-fire neuron case here, and we will assume that the neurons are exclusively in the excitation driven regime, where $I^E(t)\geq v_{thr}$ and $I^I(t)\geq v_{thr}$ for all $t>0$.  

In the limit that both $N^E\rightarrow \infty$ and $N^I \rightarrow \infty$, the population level model for the system of differential equations %in equations 
(\ref{ve1})-(\ref{vi1}) and (\ref{s}) is: 
\begin{eqnarray}
\frac{\partial \rho^E(v_e,t)}{\partial t} &=& -\frac{\partial J^E(v_e,t)}{\partial v_e} = -\frac{\partial \left(\rho^E(v_e,t)(F(v_e)+I^E(t))\right)}{\partial v_e}, \label{Fp1}\\
\frac{\partial \rho^I(v_i,t)}{\partial t} &=& -\frac{\partial J^I(v_i,t)}{\partial v_i} = -\frac{\partial\left( \rho^I(v_i,t)(F(v_i)+I^I(t))\right)}{\partial v_i}\label{Fp2}\\
\tau_E \frac{ds^E}{dt} &=&  -s^E + J^E(v_{thr},t)\\
\tau_I \frac{ds^I}{dt} &=& -s^I + J^I(v_{thr},t)
\end{eqnarray}

The Fokker-Planck system in equation (\ref{Fp1}) must be solved with the following constraints:
\begin{eqnarray*}
 J^E(v_{thr},t) = J^E(v_{reset},t), \quad J^I(v_{thr},t) = J^I(v_{reset},t), \quad \rho^E(v_e,0) = \rho^v_{0E}(v_e), \rho^I(v_i,0) = \rho^v_{0I}(v_i).
\end{eqnarray*}
This Fokker-Planck equations immediately yield the transport mean field system given by 
\begin{eqnarray*}
 \tau_E\frac{ds^E}{dt} &=& -s^E + \nu^E(t) \\
 \tau_I \frac{ds^I}{dt} &=& -s^I + \nu^I(t) \\
 I^E(t) &=& I^E_{bias} + w^{EE}s_E(t)-w^{EI}s_I(t) \\
I^I(t) &=& I^I_{bias} -w^{II}s_I(t) + w^{IE}s_E(t)\\
 \nu^E(t) &=& \rho_{0E}^v(\tilde{v}^E(t))(F(\tilde{v}^E(t))+I^E(t)), \quad \nu^I(t) = \rho_{0I}^v(\tilde{v}^I(t))(F(\tilde{v}^I(t))+I^I(t))\\
\tilde{v}^I(t) &=& g_I^{-1}\left(\text{mod}\left(-\int_0^t \Omega(I^I(t)),1\right)\right)\, \quad \tilde{v}^E(t) =g_E^{-1}\left(\text{mod}\left(-\int_0^t \Omega(I^E(t)),1\right)\right)\\
g_E(v) &=& \int_{v_{reset}}^v \frac{\Omega(I^E(t))}{F(v)+I^E(t)}, \quad g_I(v) = \int_{v_{reset}}^v \frac{\Omega(I^I(t))}{F(v)+I^I(t)} 
\end{eqnarray*}
while these equations seem formidable, they still allow for an efficient approximation to the voltage densities in the $E$/$I$ populations (Figure \ref{figure5}A-C), and approximate the firing rate fluctuations in $s^E(t)$ and $s^I(t)$ accurately over short and long time scales (Figure \ref{figure5}D-E).  We note that as in the single population case, the approximation deteriorates for long time scales (Figure \ref{figure5}E), which is to be expected in approximations of fast-slow systems. 

Our results show that solutions to the transport equation provide an alternative mean field system not based on asynchronous states that provide insights into how initial voltage densities lead to fast firing rate fluctuations and how these firing rate fluctuations evolve with weak coupling and slow inputs. 

\section*{\textbf{Discussion}}

The ``lumping" of neurons in coupled populations into neural masses has proven to be considerably difficult, but still analytically tractable, since the first phenomenological mean field systems were introduced by Wilson and Cowan \cite{cowan1}.  At the macroscale, it is not unreasonable to expect that the dynamics of many neurons should simplify somehow to a low-dimensional dynamical system which describes a network's overall qualitative and quantitative dynamics without tracking every individual neuron.    

Here, we introduced a new
%n alternative 
method to lump neurons together based on solutions to the transport equation for neurons with slowly varying inputs and weak coupling or slow synapses.
%weak coupling and/or slowly varying inputs. 
In these regimes, the transport system solution we derive acts as an approximate solution to the Fokker-Planck system, different from the asynchronous voltage density, commonly used in earlier mean field systems \cite{nesse,nicola_2d,nicola_analysis,nicola_het,nicola_noise}.  Our solution allows for the determination of both fast firing rate fluctuations and slow firing changes caused by coupling or external inputs.  The approximation leads to a marked improvement in %mimicking 
replicating the dynamics of the full network, compared with the asynchronous voltage density.  The transport system also directly links the fast firing rate fluctuations to the initial voltage density.  

We remark that considerable progress has been made in deriving mean field systems for the case of quadratic integrate-and-fire neurons with Lorentzian coupling \cite{montbrio1,montbrio2,montbrio3,montbrio4,chen1,chen2,coombes,gast1,gast2} where the Ott-Antonsen ansatz simplifies the equations governing the system dynamics considerably \cite{ott1}.  The transport mean field system considered here may also be extended to other synaptic models, neurons with heterogeneity \cite{nicola_het},  or 2-dimensional neuronal models \cite{nicola_2d,chen1}, which we leave for future work.
As in the Ott-Antonsen ansatz, it is likely that any derivation that leads to simple approximations will be heavily dependent on a chosen density function for the heterogeneity that simplifies the transport equations considerably.   The transport mean field equations are complimentary to modern neural mass models. They have a disadvantage of not being an exact reduction, and instead being an approximation based on weak-coupling/slowly varying inputs, which are usually only valid for finite time \cite{khalil}
.  One advantage the transport mean field system has is that it is %more
applicable to a whole class of neuron models.  

Another important disadvantage of the transport mean field system is that it uses the following transform to derive the solution:
%to note is that due to the transform required to go into the phase domain through $g(v)$:
$$u = g(v) = \int_{v_{reset}}^v\frac{\Omega(\bar{I}(t))dv'}{F(v')+\bar{I}(t)}. $$
This transform, and hence the transport solution, are no longer valid
when a discontinuity occurs, i.e., when $F(v) + \bar{I}(t)<0$ for some $v\in [v_{reset},v_{thr}]$.  However, the transport mean field system can be extended to accomodate this situation. In this regime, the neurons are no longer firing and their voltages tend asymptotically to a common equilibrium value (a synchronized state). %, as in this regime, some stable equilibria has emerged for neurons and they tend asymptotically to this equilibria (synchronize). 
As the flux becomes $0$ in this case, the density function must be tracked accurately to recompute the flux when $F(v)+\bar{I}(t)$ becomes positive for $v\in[v_{reset},v_{thr}]$.  This is explicitly feasible analytically for the leaky integrate-and-fire neuron, but is cumbersome as it requires re-initializing the initial voltage density after every epoch of quiescence to the tracked subthreshold voltage density.  We leave further investigation of this case for future work.

Our work shows that while the general problem of computing the mean field system for a network of interacting neurons remains difficult, considerable progress can be made if we perturb off known time-varying density solutions, rather than asynchronous ones.  By using transport-based mean field systems, an accurate theory of how initial voltage densities can influence later firing rate fluctuations can be developed.

\subsection*{\textbf{Acknowledgements}}
WN is funded by a\ Natural Sciences and Engineering Research Council Discovery Grant (DGECR-2020-00334), a Canada Research Chair (CRC-2019-00416), and through the Hotchkiss Brain Institute.  SAC is funded by a Natural Sciences and Engineering Research Council Discovery Grant (DG -2023-03611).

%\subsection*{\textbf{Code Availability Statement}}
%Accompanying code can be found at https://modeldb.science/2031430 with referee access code rxv115.  The code will be made public upon publication. 

\clearpage
\section*{\textbf{Materials and Methods}}

\subsection*{\textbf{Figure 1}}
An initial voltage density of truncated Gaussian with mean 0 and standard deviation 5 was used.  A network of $N=10^5$ leaky integrate-and-fire neurons was simulated.  A membrane time constant was set to $\tau_m = 0.2$ seconds.  The constant bias current was set to $I = -39$ pA.  All other parameters were identical to Figure 1.  The neurons were not coupled and received no time varying inputs. 

\subsection*{\textbf{Figure 2}}
The neurons consisted of $N=10^5$ LIF neurons with a bias of $I=-38$ pA.  The initial voltage densities were solved by inverting $R$ with $\rho^V =R^{-1}(z)$ with 
\begin{eqnarray}
z(t) &=& 1+\text{sign}(\sin(2\pi f\Omega(I)4t)) \quad \text{(square wave)}\\
z(t) &=& t \quad \text{(ramp)} \\ 
z(t) &=&  2+\arcsin(\sin(2\pi f(\Omega(I)5t))) \quad \text{(sawtooth wave)}\\ 
z(t) &=& \frac{1}{\sqrt{2\pi}0.02}\exp\left(-\frac{(0.2-t)^2)}{2(0.02^2)}\right) \quad \text{(Gaussian)}
\end{eqnarray}

All $z(t)$ were normalized over the interval $[0,\Omega(I)^{-1}]$ before $\rho_0(v)$ was computed with: 
\begin{eqnarray}
\rho^v{(\tilde{v})} &=& \frac{z(t^{-1}(v))}{F(\tilde{v})+I}\\
&=& R^{-1}(s) \label{eqfr1}\\
  t^{-1}(v) &=& \frac{g(v)-1}{\Omega(I)}      
\end{eqnarray}
The initial voltage densities were generated with the probability integral transform mapping uniformly drawn $u$ in $[0,1]$ to the computed target density. 

\subsection*{\textbf{Figure 3}} A network of 5000 leaky integrate-and-fire neurons was simulated with a membrane time constant $\tau_m$ of 50 milliseconds, and $v_{reset} = -60$ mV, $v_{thr} = -40$ mV.  The time constant for the filter $s(t)$ was $\tau_s=50$ ms.  The network was simulated with an integration time step of 0.01 ms, with a linear spike-time interpolant applied at each spike.  The coupling strength was $w =0.03$.  Note that all input currents/weights were also scaled by $\tau_m$.  The neurons received a constant bias current of $I = -35$ pA and were initialized with a Gaussian density with mean -50 mV and standard deviation 4 mV.  A random input current $c(t)$ was generated as
\begin{eqnarray}
    c(t) &=& A\left(\frac{\hat{c}(t) - \min(\hat{c}(t))}{\max\hat{c(t)} - \min\hat{c(t)}}\right)\\
     \hat{c}(t) &=& 15\sum_{j=1}^5 a_j \cos(2\pi f_j t)
\end{eqnarray}
where $a_j$ was drawn from a normal distribution with mean $0$ and standard deviation $1/5$, while $f_j$ was drawn from uniform distribution on $[1,2]$ Hz.   The network in Figure \ref{figure1}G-H was identical but was generated from a mixture distribution with equal mixing of two Gaussians, one with mean -54 mV, and one with a mean of -46 mV, each with a standard deviation of 1 mv.

The density function for the network was estimated with a histogram on $[v_{reset},v_{thr}]$ with intervals of 0.1 mV. An identical discretization was used to compute the transport mean field density via the equation: 
\begin{eqnarray*}
\rho^v(v,t) &=&   \rho^v_0\left(Q(v,t)\right) \left(\frac{F\left(Q(v,t)\right)+\bar{I}(t)}{F(v)+\bar{I}(t)}\right)\\
Q(v,t) &=& g^{-1}\left(\text{mod}\left(g(v) -  \int_0^t \Omega(\bar{I}(t')dt'),1\right)\right)
\label{rho1} 
\end{eqnarray*}

\subsection*{\textbf{Figure 4}}
The network was identical to that in Figure 1, only with the constant bias current of $I=-30$ pA.

\subsection*{\textbf{Figure 5}}

The network consisted of 5000 excitatory and 5000 inhibitory neurons with $w^{ee}= 0.2$, $w{ii} = 0.2$, $w^{ie} = 0.3$ and $w^{ei} = 0.1$.   The bias for the excitatory population was -30 pA while it was -20 pA for the inhibitory population. The excitatory population had a synaptic filter time constant of 50 milliseconds while the inhibitory population had a synaptic filter time constant of 100 milliseconds.  The input current to each population was drawn as in Figure 1, only with each oscillator having a randomly generated frequency in [0,2] Hz for the excitatory population, and a total amplitude of $6$ for the entire signal.  The inhibitory population had identical amplitude and frequency parameters for the input current, but with a different realization.

\clearpage

\clearpage

\clearpage

\clearpage

\clearpage

\clearpage

\bibliographystyle{unsrt}
\bibliography{references}

\end{document}